\documentclass[reprint,floatfix,noeprint,prl,superscriptaddress]{revtex4-1}
\usepackage{amsmath,amssymb,bm,braket}
\usepackage{graphicx,hyperref}
\usepackage[dvipsnames]{xcolor}

\hypersetup{colorlinks=true}
\hypersetup{citecolor=[rgb]{0.,0.,0.4}}
\hypersetup{linkcolor=[rgb]{0.,0.,0.4}}
\hypersetup{urlcolor=[rgb]{0.,0.,0.4}}
\graphicspath{{figures/}}

\allowdisplaybreaks
\newcommand{\diff}[1]{\text{d}#1\,}
\newcommand{\m}[1]{\mathcal{#1}}
\newcommand{\up}{\uparrow}
\newcommand{\down}{\downarrow}
\newcommand{\abs}[1]{\lvert#1\rvert}
\renewcommand{\vec}[1]{\bm{#1}}

\begin{document}

\title{Enhancement of Local Pairing Correlations in
Periodically Driven Mott Insulators}

\author{Francesco Peronaci}
\email{fperonaci@pks.mpg.de}
\affiliation{Université Paris-Saclay, CNRS, CEA,
Institut de Physique Théorique, 91191, Gif-sur-Yvette, France.}
\affiliation{CPHT, Ecole Polytechnique, CNRS, Universit\'{e} Paris-Saclay,
91128 Palaiseau, France}
\affiliation{Max Planck Institute for the Physics of Complex Systems,
Dresden 01187, Germany}
\author{Olivier Parcollet}
\affiliation{Université Paris-Saclay, CNRS, CEA,
Institut de Physique Théorique, 91191, Gif-sur-Yvette, France.}
\affiliation{Center for Computational Quantum Physics, Flatiron Institute,
162 Fifth Avenue, New York, NY 10010, USA}
\author{Marco Schir\'o}
\thanks{On Leave from: Institut de Physique Th\'{e}orique,
Universit\'{e} Paris Saclay, CNRS, CEA, F-91191 Gif-sur-Yvette, France}
\affiliation{JEIP, USR 3573 CNRS, Coll\`ege de France,
PSL Research  University, 11 place Marcelin Berthelot,
75231 Paris Cedex 05, France}

\date{\today}

\begin{abstract}
We investigate a model for a Mott insulator
in presence of a time-periodic modulated interaction and
a coupling to a thermal reservoir.
The combination of drive and dissipation leads to
non-equilibrium steady states with a large number of
doublon excitations, 
well above the maximum thermal-equilibrium value.
We interpret this effect
as an enhancement of local pairing correlations, providing
analytical arguments based on a Floquet Hamiltonian.
Remarkably, this Hamiltonian shows a tendency
to develop long-range staggered superconducting correlations.
This suggests the possibility of realizing
the elusive eta-pairing phase
in driven-dissipative Mott Insulators.
\end{abstract}

\maketitle

The Floquet engineering of complex quantum systems is a very active line
of research in today's condensed matter physics~\cite{Oka2018}.
It consists in the design of periodic perturbations
to achieve non-equilibrium \emph{driven} states remarkably different
from their \emph{undriven} counterparts. Examples are
the dynamical control of band topology~\cite{Oka2009,Jotzu2014}
and of magnetic interactions~\cite{Gorg2018}
in ultracold atoms in optical lattices, and of effective Hamiltonian parameters
in solids under intense laser-pulse
excitation~\cite{Kawakami2009,Singla2015,Subedi2014}.

A useful description of a periodically driven quantum system is in terms
of effective static Hamiltonians derived with large-frequency
expansions~\cite{Goldman2014,Bukov2015a,Tsuji2011,Kitamura2016}.
In general, however, the drive affects also the
distribution function of the system, eventually leading to
thermalization to a trivial infinite-temperature
state~\cite{Lazarides2014,DAlessio2014}.
Nevertheless, when heating can be avoided for finite but long times,
interesting prethermal Floquet states can be observed.
This is the case, for example, for very large
drive frequency~\cite{Abanin2015,Mori2016,Kuwahara2016,Abanin2017}
or systems close to
integrability~\cite{Bukov2015,Canovi2016,Weidinger2017,Seetharam2018,
Herrmann2018,Baum2018}.
In particular, Ref.~\cite{Peronaci2018}
showed that strong electronic correlations lead
to finite-frequency prethermal states with remarkable properties as a
function of drive frequency.

A natural question concerning the Floquet prethermal state
is whether the coupling to external reservoirs would cancel out
its interesting features,
or rather preserve them and possibly make them more accessible.
Particularly interesting is the possibility to control the distribution
function of the system by means of a dissipation mechanism of the energy
injected by the
drive~\cite{Iadecola2013,Iadecola2015,Seetharam2015,Seetharam2019,Hart2018a}.

To investigate this point, in this work we consider the Fermi-Hubbard model
with a periodically driven interaction and coupled to a thermal reservoir.
Starting from the Mott-insulating phase, our numerical calculations
show that the combination of drive and dissipation leads to
steady states that are not accessible in the corresponding isolated model.
In particular, we reveal a regime with a remarkably large number of high-energy
doublon excitations,
well above the maximum equilibrium value for the half-filled
repulsive Hubbard model.

We interpret this steady-state large double occupancy as
an enhancement of local pairing correlations,
and we describe the effect as a thermalization
to a lowest-order Floquet Hamiltonian.
Remarkably, we find that higher-order terms promote
finite-momentum doublon superfluidity,
namely staggered long-range pairing correlations among fermions
($\eta$-pairing),
which spontaneously break the hidden SU$_\text{C}(2)$ charge symmetry
of the half-filled Hubbard model~\cite{Yang1989,Zhang1990,Mitra2018}. 
This suggests a nonequilibrium protocol
for Floquet engineering exotic superconducting
states in driven-dissipative Mott insulators,
as also very recently investigated in similar
contexts~\cite{Kaneko2019,Tindall2019,Werner2019,Li2019}.

Our results are relevant for current experiments on
laser-pumped organic Mott
insulators~\cite{Kawakami2009,Singla2015} and ultracold Fermi gases in driven
optical lattices~\cite{Messer2018,Sandholzer2019}.
We discuss the latter in particular, suggesting to explore a possibly
overlooked regime in future experiments.

\emph{Model --}
The Hamiltonian of the driven-dissipative Fermi-Hubbard model reads
$H=H_\text{Hub}+H_\text{diss}$, where:
\begin{gather}
\label{eq_1}
H_\text{Hub}= \sum_{ij,\sigma}V_{ij}c^\dagger_{i\sigma}c_{j\sigma}
+U(t)\sum_i(n_{i\up}-\tfrac{1}{2})(n_{i\down}-\tfrac{1}{2}),\\
\label{eq_2}
H_\text{diss}=
\sum_{i\alpha}\omega_\alpha b^\dagger_{i\alpha}b_{i\alpha}
+\lambda\sum_{i\alpha}g_{\alpha}(n_{i}-1)(b_{i\alpha}+b_{i\alpha}^\dagger).
\end{gather}
Here the $c$'s operators describe fermions hopping with amplitude $V_{ij}$
and subject to a driven local interaction
$U(t\ge0)=U_0+\delta U\sin\Omega t$.
The bare density of states is semicircular with bandwidth $4V$ and
we measure energy, frequency and inverse of time ($\hbar=1$) in units
of $V$~\cite{Note2}.
The thermal bath is implemented by independent sets of bosonic modes $b$'s
which couple to density at each lattice site, with spectral function
$J(\omega)=\sum_\alpha g^2_\alpha\delta(\omega-\omega_\alpha)
\propto\omega^2e^{-\frac{\omega}{\omega_c}}$ ($\omega_c=1$)
and coupling $\lambda$~\cite{Note2}.
Importantly, the bath allows energy dissipation but commutes
with density $n_i=n_{i\uparrow}+n_{i\downarrow}$
and preserves particle-hole
symmetry. The system remains half filled
at all times ($\braket{n_{i\sigma}}=0.5$).

\begin{figure*}
\includegraphics[width=\textwidth]{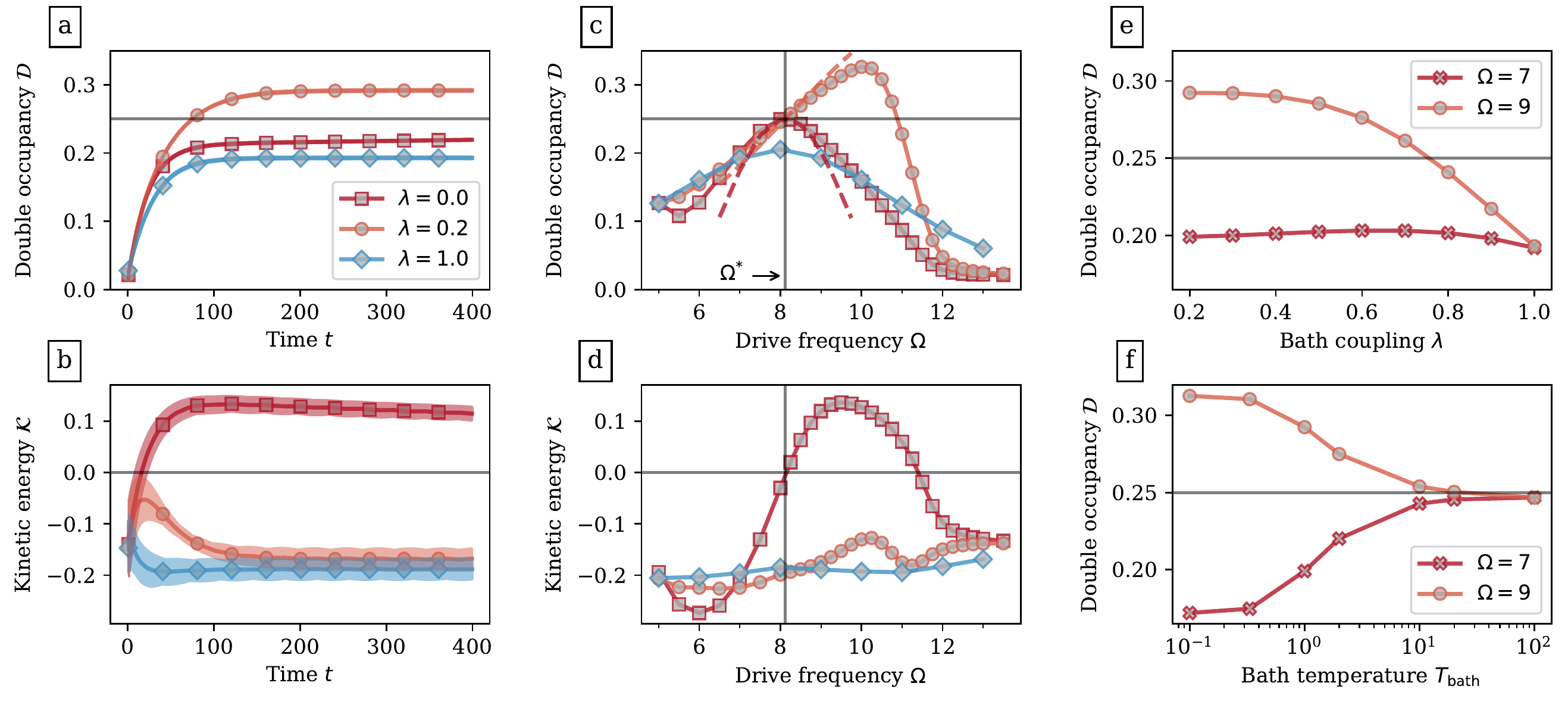}
\caption{(a-b): Time evolution of double occupancy and kinetic energy
for drive frequency $\Omega=9>\Omega^*$:
oscillations (shaded area) and their average (symbols).
Legend in (a) is common to (a-b-c-d). (c-d): Long-time averages
as a function of drive frequency, and approximate analytical expressions
based on Eq.~\eqref{eq_4} (dashed lines).
(e-f): Long-time average of double occupancy
as a function of bath coupling and of bath temperature for $\lambda=0.2$
and drive frequencies $\Omega=7<\Omega^*$ and $\Omega=9>\Omega^*$.}
\label{fig_1}
\end{figure*}

Starting from a thermal-equilibrium state, 
we calculate the time evolution by means of nonequilibrium dynamical
mean-field theory (DMFT)~\cite{Georges1996,Aoki2014} with the
non-crossing approximation as impurity solver~\cite{Eckstein2010,Ruegg2013},
including the effect of dissipation at weak coupling in
$\lambda$~\cite{Chen2016}
(see Supplemental Material~\cite{Note1} Sec.~I for implementation details,
and Sec.~II for one-crossing-approximation benchmarks).
We calculate double occupancy
$\mathcal{D}(t)=\braket{n_{i\up}(t)n_{i\down}(t)}$,
kinetic energy
$\mathcal{K}(t)=\sum V_{ij}\braket{c^\dagger_{i\sigma}(t)c_{j\sigma}(t)}$,
and local Green's function
$G_\sigma(t,t')=-i\braket{\text{T} c_{i\sigma}(t)c^\dagger_{i\sigma}(t')}$.
For definiteness, we choose $U_0=8$
for the initial Mott-insulating state
in equilibrium at $T=1$~\cite{Note3}
and $\delta U=2$ for the drive amplitude
(see Supplemental Material~\cite{Note1} Sec.~VI for a discussion
on $\delta U$). The bath temperature is $T_\text{bath}=1$
unless specified differently.
In absence of dissipation, Floquet prethermalization is observed at all
frequencies except for the resonance
$\Omega^*=8.12\simeq U_0$~\cite{Peronaci2018}.
We restrict ourselves to paramagnetic states,
leaving the interplay of drive, dissipation and magnetism
to future studies.

\emph{Time evolution --} In the driven-dissipative model, as well as in the
isolated case, double occupancy and kinetic energy display a separation of
time scales between fast oscillations synchronized with the drive and a
slowly varying average value.
However, after a common transient,
the thermal reservoir starts to be effective and changes substantially
the long time behavior of both observables.

For weak bath coupling and drive above resonance,
the double occupancy grows substantially larger than in the isolated model,
going to a stationary average above $0.25$
(Fig.~\ref{fig_1}a, $\lambda=0.2$).
Such a large value would be possible, at equilibrium, only
if the interaction were \emph{attractive}. This striking effect
highlights the peculiarity of this non-equilibrium steady
state, as we discuss thoroughly below.

Upon increasing the bath coupling (Fig.\,\ref{fig_1}a, $\lambda=1.0$),
the double occupancy decreases
and eventually remains below the limit of $0.25$ at all times.
Moreover, we notice that the bath is effective only after a transient
time $\sim 1/\lambda^2$, which makes the regime of very weak
coupling not accessible by the numerical simulation
(see also Ref.~\cite{Eckstein2013}).

At the same time, the kinetic energy is also largely affected
by dissipation (Fig.\,\ref{fig_1}b). Here the effect is more intuitive:
in the isolated model the drive leads to a prethermal state with positive
kinetic energy, indicative of a population
inversion~\cite{Herrmann2018,Peronaci2018}. On the other hand, the thermal
reservoir dissipates the excess kinetic energy, which remains negative as at
equilibrium, and inhibits the population inversion, as we also explicitly
show below.

\emph{Long-time average --} 
To study the role of drive frequency and bath coupling in a more systematic way,
we consider the long-time average of double occupancy and kinetic energy.
For weak bath coupling, the dissipative model has double occupancy
larger than the isolated one at all frequencies
(Fig.\,\ref{fig_1}c, $\lambda=0.2$).
However, a remarkable change happens crossing the resonance
$\Omega^*\simeq U_0$. Below resonance, the dissipation has only a
quantitative, rather weak effect. In contrast, above resonance,
we systematically observe a large increase of double occupancy across
the limit of $0.25$, as discussed previously for a selected frequency.
Lower values are then recovered upon increasing the frequency further,
as the system eventually becomes transparent to the drive.

Independent of the bath coupling, the kinetic energy of the dissipative model
is rather featureless and negative for all frequencies (Fig.\,\ref{fig_1}d,
$\lambda=0.2,1.0$).
Thus, the thermal reservoir cancels the region of positive kinetic energy
characteristic of the isolated case (Fig.\,\ref{fig_1}d, $\lambda=0.0$).

The difference between below and above resonance appears also in the
dependence on bath coupling (Fig.\,\ref{fig_1}e) and bath temperature
(Fig.\,\ref{fig_1}f). Below resonance ($\Omega=7$)
the double occupancy is almost independent of bath coupling
and decreases on lowering
the bath temperature. Quite differently, above resonance ($\Omega=9$)
it increases on lowering the bath temperature at weak bath coupling.

\begin{figure}
\includegraphics[width=\columnwidth]{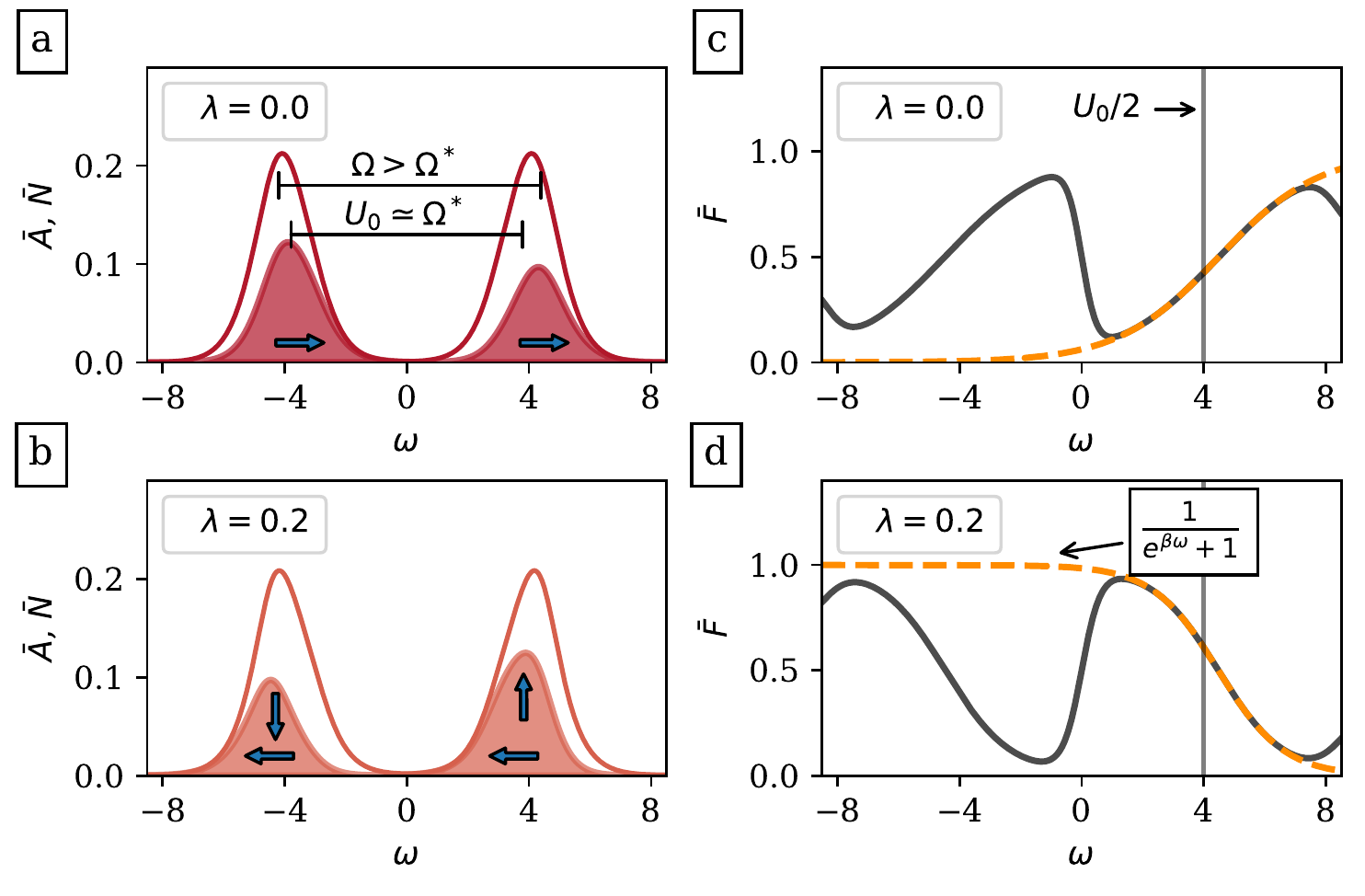}
\caption{
(a-b): Long-time average spectral function $\bar A(\omega)$ (solid line)
 and occupation function $\bar N(\omega)$ (filled area)
of the isolated model ($\lambda=0$) and the dissipative model ($\lambda=0.2$)
for drive frequency $\Omega=9>\Omega^*$. 
The isolated model (a) has population inversion,
signaled by the blueshift of
$\bar N(\omega)$ with respect to $\bar A(\omega)$ (arrows).
In the dissipative model (b),
the thermal reservoir cancels the population inversion (horizontal arrows)
and unveils a non-thermal state with large double occupancy,
signaled by the increase of $\bar N(\omega)$ in the high-energy band
(vertical arrows). (c-d): Long-time average distribution
function $\bar F(\omega)$ for the same
parameters of (a-b)
with Fermi-function fit around the Hubbard-band center.
The extracted effective temperature is $T_\text{eff}=-1.6$
for the isolated model (c) and $T_\text{eff}=1.1\simeq T_\text{bath}$
for the dissipative one (d).
See Supplemental Material~\cite{Note1} Fig.~S3 for a plot of
$T_\text{eff}(\Omega)$.}
\label{fig_2}
\end{figure}

Notice that the observed behavior does not depend on the details of
the bath spectral function as long as  we deal with bosonic modes~\cite{Note2}.
In contrast, a fermionic reservoir does not lead to the same steady-state
large double occupancy (see Supplemental Material~\cite{Note1} Sec.~III)
because it can change the local density even at zero hopping,
spoiling the quasi-conservation of doublons which is the basis
of Floquet prethermalization in this system~\cite{Peronaci2018}.

\emph{Spectral function --} To gain insight into
the nature of the steady state, we calculate the 
spectral function $\bar A(\omega)$ and occupation
function $\bar N(\omega)$ as the average Wigner transforms of
the retarded and lesser components of the
local Green's function~\cite{Peronaci2018}.
While the spectral function is the same in the isolated and dissipative models,
the occupation function, and thus the distribution function
$\bar F(\omega)=\bar N(\omega)/\bar A(\omega)$,
changes drastically for drive frequency
above resonance and weak bath coupling.

In the isolated model, $\bar N(\omega)$ is shifted towards high energy
with respect to $\bar A(\omega)$ (Fig.~\ref{fig_2}a).
There is therefore a population
inversion within the Hubbard bands. Indeed,
the local behavior of $\bar F(\omega)$ for $\omega\simeq\pm U_0/2$
 has the shape of a Fermi function
with negative temperature (Fig.~\ref{fig_2}c).

The thermal reservoir completely changes the situation.
First, as the dissipation enhances the
energy redistribution within the Hubbard bands,
$\bar N(\omega)$ is pushed back to lower energy (Fig.~\ref{fig_2}b),
cancelling the population inversion.
As a consequence, $\bar F(\omega)$ assumes the shape of a
Fermi function with positive
temperature for $\omega\simeq\pm U_0/2$ (Fig.~\ref{fig_2}d).
Then, the overall weight of $\bar N(\omega)$ in the upper band grows and
becomes even larger than in the lower band, meaning the creation of a large
number of high-energy doublon excitations.
These two effects are qualitatively related to the ones discussed above:
change of sign of kinetic energy and growth of double occupancy.

\emph{Discussion --}
The above numerical results demonstrate that,
in the strongly repulsive Fermi-Hubbard model, the combination of 
a time-periodic interaction and a dissipative bath leads to steady states
with a remarkably large number of doublon excitations.
Interestingly, this large double occupancy
immediately translates into enhanced local pairing correlations
$\m{D}=\braket{c_{i\uparrow}^\dagger c_{i\downarrow}^\dagger
c_{j\downarrow} c_{j\uparrow}}|_{i=j}$, although a full calculation 
of the lattice susceptibility within DMFT is beyond the scope of this paper.

In order to unveil the origin of this effect, let us first consider the
isolated model and its Floquet Hamiltonian. To this end, we consider a
frequency close to resonance $\Omega\simeq\Omega^*\simeq U_0$ and
perform a rotating-frame transformation on the Hamiltonian~\eqref{eq_1},
followed by a high-frequency expansion~\cite{Bukov2015a,Bukov2016}
(see Supplemental Material~\cite{Note1} Sec.~V).
At lowest order we find:
\begin{equation}
\label{eq_4}
\bar H_\text{Hub}^{\mathrm{eff}(0)} = VK_0
+(U_0-\Omega)\sum\nolimits_i(n_{i\uparrow}-\tfrac{1}{2})
(n_{i\downarrow}-\tfrac{1}{2}).
\end{equation}
Here
$K_0=\sum(V_{ij}/V)c_{i\sigma}^\dagger c_{j\sigma}
(n_{i\bar\sigma}n_{j\bar\sigma}+\bar n_{i\bar\sigma}\bar n_{j\bar\sigma})$
are those hopping terms in Eq.~\eqref{eq_1} 
that do not alter the number of doubly occupied sites
($\bar n_{i\sigma}=1-n_{i\sigma}$, $\bar\uparrow=\downarrow$
and $\bar\downarrow=\uparrow$). 
Eq.~\eqref{eq_4} can be interpreted as
a Hamiltonian of doublons and holons, where
the first term contains hopping
and the second term acts as a chemical potential.
The numerical results around $\Omega^*\simeq U_0$
in the isolated model are qualitatively captured in terms of thermalization
to this effective Hamiltonian.
Indeed, we can extract the effective temperature
$T_\text{eff}(\Omega) \simeq (U_0-\Omega)^{-1}$
(see Supplemental Material~\cite{Note1} Fig.~S3) and,
since $\abs{T_\mathrm{eff}}\gg V$ for $\Omega\simeq\Omega^*$,
we can disregard the kinetic term in Eq.~\eqref{eq_4} and calculate
$\m{D}=0.5[1+\exp(0.5 (U_0-\Omega)/T_\text{eff})]^{-1}
\simeq 0.25-(\Omega-U_0)^2$. This captures
the quadratic behavior for $\lambda=0.0$ (Fig.~\ref{fig_1}c; dashed line)
with finite-hopping corrections responsible for the quantitative mismatch.

We now turn to the dissipative model. Here two observations are crucial.
First, the enhancement of double occupancy
is most pronounced for weak bath coupling (see Fig.\,\ref{fig_1}e).
Second, the dissipation leaves largely unchanged the spectral function
of the system, while it profoundly changes its occupation
(see Fig.\,\ref{fig_2}).
On this basis, we argue that at weak coupling the bath does not
change the Floquet Hamiltonian,
but only affects the effective temperature
$T_\text{eff}\simeq T_\text{bath}=1$
(see Supplemental Material~\cite{Note1} Fig.~S3).
This leads to $\m{D}\simeq0.25+(\Omega-U_0)$
which qualitatively reproduces the behavior
for $\lambda=0.2$ around $\Omega\simeq\Omega^*$
(Fig.\,\ref{fig_1}c; dashed line) where again the mismatch is
due to the finite hopping in Eq.~\eqref{eq_4}.

The outcome of this analysis is that, at least for
moderately high temperatures $T_\text{bath}\simeq V$,
the driven-dissipative protocol of Eqs.~\eqref{eq_1} and \eqref{eq_2}
leads at long times to thermal states of the doublon Hamiltonian~\eqref{eq_4}.
Singly occupied sites, which are relevant in the transient dynamics
during doublon-holon proliferation, are not relevant for the
steady-state physics, and can be considered as a reservoir of energy
and particles to the doublon-holon system.

A natural question now is whether the enhanced local pairing 
correlations can propagate through the lattice giving a superfluid state
of doublons. To answer this, we consider the next order in the
Floquet Hamiltonian~\cite{Note1}:
\begin{equation}
\label{eq_5}
\bar H_\text{Hub}^{\mathrm{eff}(1)} =
(-i\m{J}_{1}(\tfrac{\delta U}{\Omega})VK_+
+ \text{H.c.})
+\tfrac{V^2}{\Omega}(\m{J}_0(\tfrac{\delta U}{\Omega}))^2[K_+,K_-].
\end{equation}
Here $\m{J}_n(x)$ is the $n$-th order Bessel function of the first kind,
$K_+=\sum(V_{ij}/V) c_{i\sigma}^\dagger c_{j\sigma}
n_{i\bar\sigma}\bar n_{j\bar\sigma}=(K_-)^\dagger$ and one has to note that,
in the case of weak drive amplitude considered here,
$\m{J}_n(\delta U/\Omega)\sim(\delta U/\Omega)^n$ and therefore all terms
in Eq.~\eqref{eq_5} indeed vanish as the inverse drive frequency.

The first two terms in parentheses in Eq.~\eqref{eq_5}  
create or annihilate doublon excitations, controlling
the transient dynamics.
However, these processes are largely inhibited in the steady state.
Indeed, these terms depend strongly on the drive
amplitude $\delta U$, which controls the transient time-scale
but does not influence the long-time steady-state, as found
in both the isolated~\cite{Peronaci2018} and the dissipative models
(see Supplementary Material~\cite{Note1} Sec.~V).

The last term in Eq.~\eqref{eq_5} is similar to the
Schrieffer-Wolff result~\cite{Bukov2016,MacDonald1988},
which is retrieved for $\delta U\rightarrow0$
at fixed $\Omega=U_0$. It contains several contributions
such as density and exchange interactions, and correlated
three-site hopping processes.
At equilibrium and half filling, in the relevant limit of zero double
occupancy, this gives the usual anti-ferromagnetic Heisenberg model.
In far-from-equilibrium situations, if a sizeable population of doublons
is achieved as in the present case, the same term leads to a completely
different physics (see also Ref.~\cite{Rosch2008}).

To discuss Hamiltonian~\eqref{eq_5} on states with very large double occupancy,
it is instructive to neglect processes involving singly
occupied sites and rewrite it as~\cite{Note1}:
\begin{equation}
\label{eq_6}
\bar H_\text{Hub}^{\mathrm{eff}}
= J_\text{eff}
\sum\nolimits_{\braket{ij}}(c_{i\uparrow}^\dagger c_{i\downarrow}^\dagger
c_{j\downarrow} c_{j\uparrow}
+n_{i\uparrow}n_{i\downarrow}\bar n_{j\uparrow}\bar n_{j\downarrow}).
\end{equation}
Here $J_\text{eff}=2V^2/\Omega(\m{J}_0(\delta U/\Omega))^2$,
the first term is a doublon hopping,
and the second term a first-neighbor doublon interaction.
It is now convenient to consider a transformation on spin-down operators
$c_{i\downarrow}\rightarrow\tilde c_{i\downarrow}=(-1)^i c_{i\down}^\dagger$
which recasts Eq.~\eqref{eq_6} as 
$\bar H_\text{Hub}^{\mathrm{eff}}=
-J_\text{eff}\sum\vec\eta_i\cdot\vec\eta_j$ namely as an
isotropic ferromagnetic
Heisenberg model
for the so-called
$\eta$-spins
$\vec{\eta}_i=\frac{1}{2}\sum_{\alpha\beta}
\tilde c_{i\alpha}^\dagger\vec\sigma_{\alpha\beta}
\tilde c_{i\beta}$~\cite{Note1}.
The invariance under $\eta$-spin rotation is associated
to the charge SU$_\text{C}$(2) symmetry of
Hamiltonian~\eqref{eq_1} which
can be used to build eigenstates of the Hubbard model
with staggered long-range superconducting correlations
($\eta$-pairing)~\cite{Yang1989,Zhang1990}.
These same correlations are encoded in
Eq.~\eqref{eq_6}.
Indeed, the $\eta$-spin ferromagnetic Heisenberg model
has magnetization  $\braket{\eta^z} = \m{D}-0.5$
and below a critical temperature $\sim J_\text{eff}$ it has
finite order parameter in the $xy$-plane,
which corresponds to
staggered long-range pairing correlations
$\braket{c^{\dagger}_{i\uparrow}c^{\dagger}_{i\downarrow}
+\text{H.c.}}=(-1)^i\braket{2\eta_i^x}= (-1)^i
\sqrt{4\m{D}(1-\m{D})}$.

We stress again that here, for simplicity,
we do not consider the interplay between doublons and
singly occupied sites, which would result in additional terms
in Eq.~\eqref{eq_6}.
Moreover, we notice that the SU$_\text{C}(2)$ symmetry implies
a degeneracy between the $xy$-plane
and the $z$-axis of the $\eta$-spin, which translates into a competition
between superfluidity and charge-density wave.
We leave the investigation of these issues for future work.

The model system investigated here can be realized in current
experimental platforms.
Particularly promising are Mott-insulating organic molecular crystals,
where laser excitations can
induce an effective time-periodic modulation
of interaction~\cite{Kawakami2009,Singla2015}.
More direct control is achieved with ultracold atoms in optical lattices.
Recent experiments~\cite{Messer2018,Sandholzer2019} have studied the Floquet
prethermal state and remarkably
found large double occupancy for drive above resonance.
We suggest that also in these cases a key role
is played by dissipation, which is unavoidable even in cold atoms.
Finally, we notice that these experiments have focused
on the regime where the effective Hamiltonian
reduces to a renormalized Hubbard model. In contrast, here we have studied 
the case of a doublon-only Floquet Hamiltonian.
Therefore, we suggest future experiments to investigate this latter regime
to detect the presence of staggered pairing correlations.

\emph{Conclusions --}
In this work, we have studied the combination of a periodically
driven interaction and a dissipative bath in the
strongly repulsive Fermi-Hubbard model.
For weak bath coupling and frequency in a range above the
resonance of the isolated model~\cite{Peronaci2018},
we find a large increase of double occupancy,
well above the maximum equilibrium value, which
we interpret as an enhancement of local pairing correlations,
and understand in terms of
thermalization to the lowest-order Floquet Hamiltonian.

Remarkably, the next-order Floquet Hamiltonian contains terms
which promote staggered pairing correlations.
Therefore, provided a nonequilibrium protocol
to reach low enough effective temperatures
(see e.g.\ Ref.\,\cite{Werner2019b})
and eventually further increase the doublon density,
the steady state of the driven-dissipative
Fermi-Hubbard model would contain off-diagonal staggered long-range order
($\eta$-pairing),
hence a superfluid phase of doublon excitations, similarly to what very
recently found in similar models~\cite{Werner2019,Li2019}.

\begin{acknowledgments}
This work was supported by the European Research Council grants
ERC-319286-QMAC (FP) and ERC-278472-MottMetals (FP, OP).
The Flatiron Institute is a division of the Simons Foundation.
MS acknowledges support from a grant ``Investissements d'Avenir''
from LabEx PALM (ANR-10-LABX-0039-PALM)
and from the CNRS through the PICS-USA-14750.
\end{acknowledgments}

\bibliography{main}

\footnotetext[1]{See Supplemental Material at [url] for details on the
numerical implementation, benchmarks with one-crossing-approximation,
effective temperature analysis, derivation of the Floquet Hamiltonian,
and additional data with fermionic bath and with different drive amplitude,
lattice bare density of states and bath spectral function.}
\footnotetext[2]{
We expect our results to hold also for other lattices and
bath parameters, as the key ingredients --  strong local correlations
and energy dissipation -- do not significantly depend on these
details. In Supplemental Material~\cite{Note1} Secs.~VII-VIII we
provide additional data on hypercubic lattice and with different bath
parameters.}
\footnotetext[3]{Our results do not qualitatively depend on $U_0$ and $T$,
as long as the initial state is in the Mott-insulating phase.}

\clearpage

\setcounter{equation}{0}
\renewcommand{\theequation}{S\arabic{equation}}
\setcounter{figure}{0}
\renewcommand{\thefigure}{S\arabic{figure}}

\subsection*{Supplemental Material}

\section{I. NCA with bosonic bath}
Here we describe our implementation of 
a bosonic bath
in the non-crossing-approximation (NCA) impurity solver for
non-equilibrium dynamical mean-field theory (DMFT).
Starting from Hamiltonians~\eqref{eq_1} and \eqref{eq_2}
in main text, we integrate out the bosons and obtain
the action:
\begin{equation}
\small
\label{eq_s1}
\begin{split}
S_\text{latt}
&=\int\diff{t} \sum_{ij,\sigma}V_{ij} c^\dagger_{i\sigma}(t)c_{j\sigma}(t)\\
&+\int\diff{t}
U(t)\sum_i(n_{i\up}(t)-\tfrac{1}{2})(n_{i\down}(t)-\tfrac{1}{2})\\
&+\lambda^2\int\int\diff{t}\diff{t'}\Delta_b(t,t')\sum_i
(n_i(t)-1)(n_i(t')-1).
\end{split}
\end{equation}
Here the integrals are along the three-branch Keldysh contour and
the bath enters via the hybridization
$\Delta_b(t,t') = -i\int\diff{\omega}J(\omega)
(\theta(t,t')+n_B(\omega/T_\text{bath})) e^{-i\omega(t-t')}$ where
$\theta(t,t')$ is the Heaviside theta function on contour
and $n_b(\omega)$ is the Bose distribution.

In DMFT the lattice action~\eqref{eq_s1} is mapped onto the
action of a quantum impurity  coupled to
a self-consistent fermionic bath:
\begin{equation}
\label{eq_s2}
\small
\begin{split}
S_\text{imp}
&=\int\diff{t}U(t)(n_\up(t)-\tfrac{1}{2})(n_\down(t)-\tfrac{1}{2})\\
&+V^2\int\int\diff{t}\diff{t'}G(t,t')
\sum_\sigma c_\sigma^\dagger(t)c_\sigma(t')\\
&+\lambda^2\int\int\diff{t}\diff{t'}\Delta_b(t,t')(n(t)-1)(n(t')-1).
\end{split}
\end{equation}
Here we have used the 
relation $\Delta(t,t')=V^2G(t,t')$
for the hybridization $\Delta(t,t')$ of the self-consistent fermionic bath,
which is valid on Bethe lattice.

To derive the NCA equations,
we expand the partition function $\text{Tr}(\exp[-iS_\text{imp}])$
into a power series in $V$ and $\lambda$ and truncate the expansion at
the first self-consistent
order~\cite{Eckstein2010,Ruegg2013,Chen2016}.
This series is
expressed in terms of the
propagator of the states of the impurity $R$, and of its
self-energy $S$, which satisfy an integro-differential
equation similar to the usual Dyson
equation (see Ref.~\cite{Peronaci2018} for our implementation).
Then, in the present case a convenient basis choice is
$\{\ket{0},\ket{\up},\ket{\down},\ket{\up\down}\}$ which makes the propagator
$R$ and the self-energy $S$ diagonal.
Finally, we exploit the spin SU$(2)$ symmetry and the particle-hole 
SU$_\text{C}(2)$ symmetry of Eq.~\eqref{eq_s2} which further reduce the 
number of propagators to two: one for the empty state $\ket{0}$ and one for 
the singly-occupied state $\ket{\uparrow}$,
with the following self-energies:
\begin{align}
\small
S_{\ket{0}}(t,t')=&-2iR_{\ket{\uparrow}}(t,t')\Delta(t',t)\\
&+i\lambda^2 R_{\ket{0}}(t,t')(\Delta_b(t,t')+\Delta_b(t',t)),\nonumber\\
S_{\ket{\uparrow}}(t,t')=&+2iR_{\ket{0}}(t,t')\Delta(t,t').
\end{align}

\section*{II. OCA benchmark}
In this work we investigate the Mott-insulating phase of the
Fermi-Hubbard model with
average interaction parameter $U_0=8V$ much larger than
the bare band-width $W=4V$. Moreover, we consider initial thermal
density matrix at a rather high temperature $T=1$. In these conditions,
the NCA approximation is expected to perform well both at equilibrium
and out of equilibrium~\cite{Eckstein2010}.

\begin{figure}
\includegraphics[width=\columnwidth]{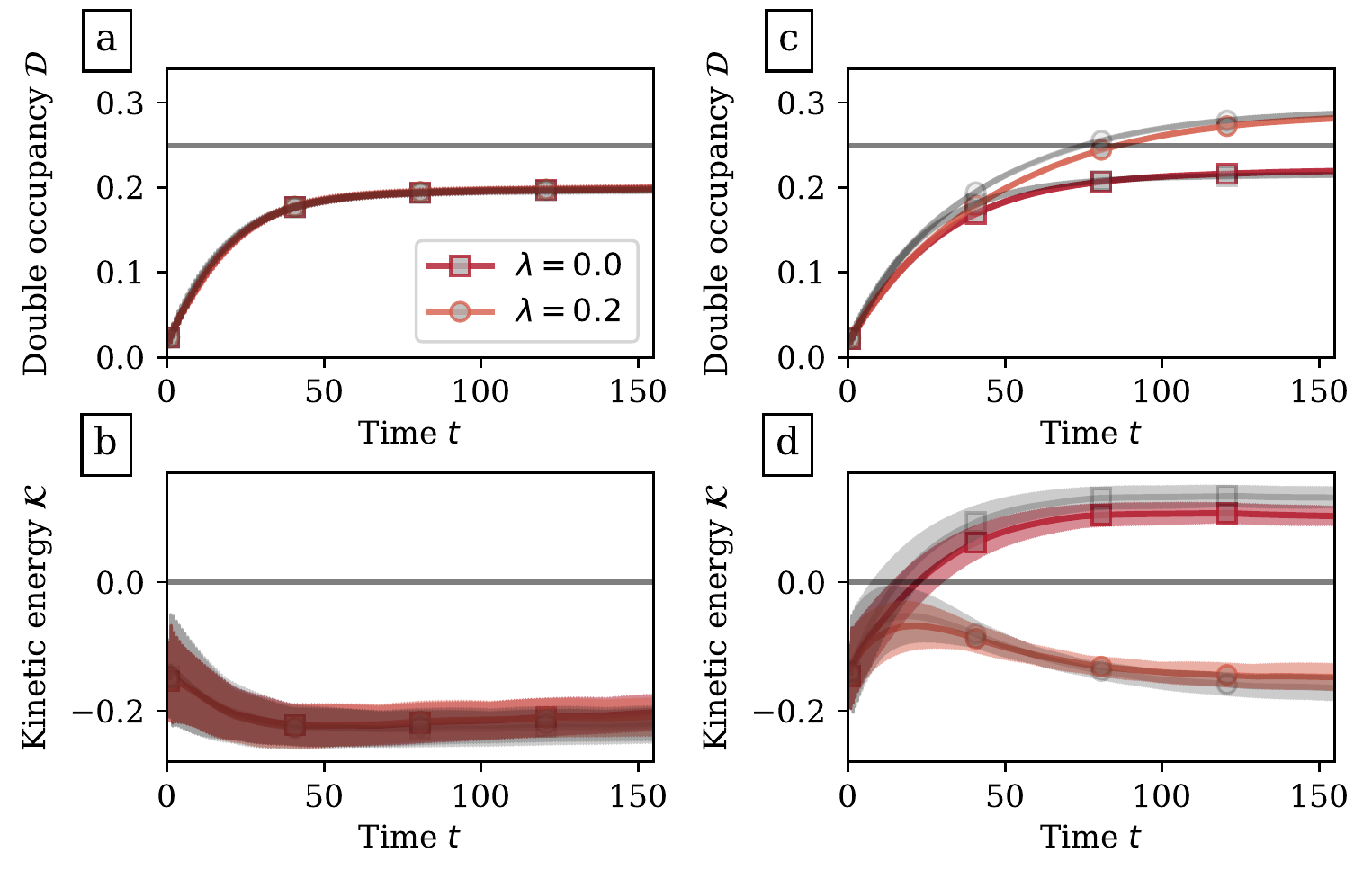}
\caption{Time evolution of double occupancy and kinetic energy
using the OCA impurity solver for DMFT: Isolated model ($\lambda=0$)
and dissipative model ($\lambda=0.2$) for drive frequency below
resonance $\Omega=7$ (a-b) and above resonance $\Omega=9$ (c-d).
Legend in (a) is common to (a-b-c-d). Light grey curves are the corresponding
NCA results, for comparison.}
\label{fig_sm_3}
\end{figure}

To confirm this, we have performed some calculations using the next-order
one-crossing approximation (OCA).
This takes into consideration terms of order $\m{O}(V^4)$ in the
hybridization expansion~\cite{Eckstein2010,Ruegg2013}
(see Ref.~\cite{Peronaci2018} for our implementation).
As expected, the time evolution of double occupancy and kinetic energy
(Fig.~\ref{fig_sm_3}) is essentially identical to the one shown
in the main text (Fig.~\ref{fig_1}a-b).

\section*{III. Fermionic bath}
It is interesting to consider an external reservoir of fermionic modes,
instead of the bosonic bath considered in the main text.
To do so, we substitute Eq.~\eqref{eq_2} of the main text with the following
coupling to a fermionic bath:
\begin{equation}
\small
H_\text{diss}=
\sum_{i\alpha}\omega_\alpha f^\dagger_{i\alpha}f_{i\alpha}
+\lambda\sum_{i\sigma\alpha}(g_{\alpha}c_{i\sigma}^\dagger f_{i\alpha}+
g_{\alpha}^*f_{i\alpha}^\dagger c_{i\sigma}).
\end{equation}
Here the operators $f$'s represent sets of independent fermionic harmonic
oscillators. In this case, the system can dissipate both energy
and particles. Moreover, this type of coupling can be treated exactly: 
integrating out the fermionic bath, we 
introduce an additional hybridization
$\Delta_f(t,t')$ which can simply be added to the DMFT self-consistent
hybridization $\Delta(t,t')$.

\begin{figure}
\includegraphics[width=\columnwidth]{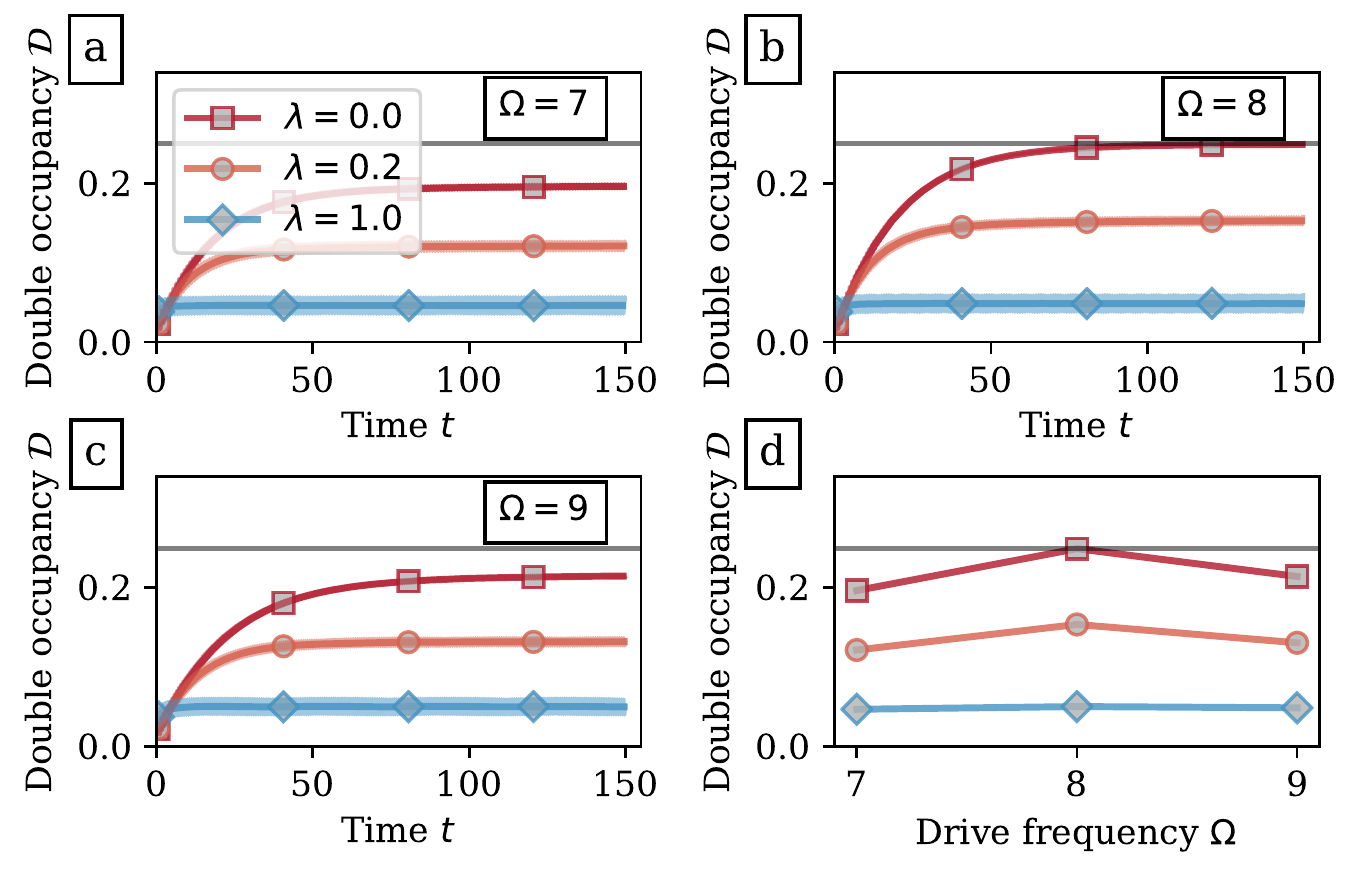}
\caption{Panels (a-b-c): Time evolution of double occupancy
for the isolated model ($\lambda=0.0$) and the dissipative one with
fermionic bath ($\lambda=0.2,1.0$) for various drive frequencies.
Legend in (a) is common to (a-b-c-d).
Panel (d): Long-time average of double
occupancy as a function of drive frequency.}
\label{fig_sm_4}
\end{figure}

As shown in Fig.~\ref{fig_sm_4}, the coupling with a fermionic bath
does not lead to the same interesting features discussed
in the main text for the bosonic bath (cf. Fig.~\ref{fig_1}a-c).
This is due to the fact that the bosonic bath,
 as opposed to the fermionic bath,
 commutes with the local density and conserves double occupancy,
and therefore it preserves the mechanism for
Floquet prethermalization.

\section*{IV. Effective temperature}
In Fig.~\ref{fig_2}
of main text we show the steady-state average distribution function
$\bar F(\omega)$  for $\Omega=9>\Omega^*$, along with a
Fermi-function fit $[1+\exp(\omega/T_\text{eff})]^{-1}$ around the
center of the upper Hubbard band $\omega\simeq U_0/2$.
From this fit we can extract the effective temperature $T_\text{eff}$.
Here, in Fig.~\ref{fig_sm_4} we plot this effective temperature 
for the isolated and the dissipative models, as a function of the drive
frequency.

\begin{figure}
\includegraphics[width=\columnwidth]{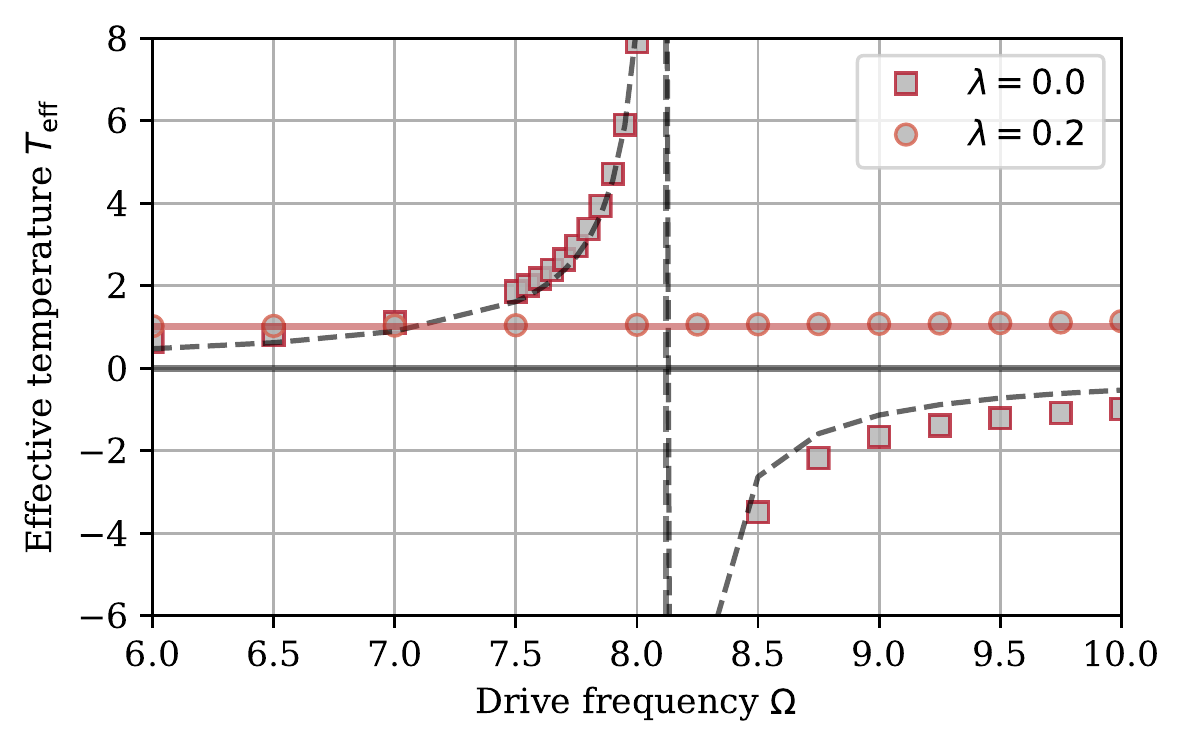}
\caption{Effective temperature (symbols) as a function
of drive frequency for the isolated model ($\lambda=0.0$)
and the dissipative one ($\lambda=0.2$). In the former, the
effective temperature diverges as $(\Omega^*-\Omega)^{-1}$ (dashed line).
The divergence at $\Omega=\Omega^*$ signals
the resonant thermalization,
while the negative effective temperature for $\Omega>\Omega^*$
signals the population inversion~\cite{Peronaci2018}.
In contrast, the dissipative model has temperature $\simeq T_\text{bath}$
(solid line) fixed by the thermal reservoir.}
\label{fig_sm_5}
\end{figure}

\section*{V. Rotating-frame transformation and high-frequency expansion}
To carry out the large-frequency expansion~\cite{Bukov2015a,Bukov2016},
we first need to transform Hamiltonian~\eqref{eq_1} of main text
to a rotating frame with respect to the interaction:
\small
\begin{align}
H_\text{Hub}(t)&=\sum V_{ij}c^\dagger_{i\sigma}c_{j\sigma}
+U(t)\sum\nolimits_i(n_{i\up}-\tfrac{1}{2})(n_{i\down}-\tfrac{1}{2}),\\
\bar H_\text{Hub}(t) &= e^{S(t)}(H_\text{Hub}(t)-i\partial_t)e^{-S(t)},\\
S(t)&=iF(t)\sum\nolimits_i(n_{i\up}-\tfrac{1}{2})(n_{i\down}-\tfrac{1}{2})
=iF(t)D,\\
F(t)&=\Omega t -(\delta U/\Omega)\cos\Omega t.
\end{align}
\normalsize
It is useful to introduce the operators $K_0$ and $K_\pm$ as in the main text
($\bar n_{i\sigma}=1-n_{i\sigma}$, $\bar\uparrow=\downarrow$
and $\bar\downarrow=\uparrow$):
\small
\begin{align}
K_0&=\sum\nolimits_{ij\sigma}(V_{ij}/V)c_{i\sigma}^\dagger c_{j\sigma}
(n_{i\bar\sigma}n_{j\bar\sigma}+\bar n_{i\bar\sigma}\bar n_{j\bar\sigma}),\\
K_+&=\sum\nolimits_{ij\sigma}(V_{ij}/V) c_{i\sigma}^\dagger c_{j\sigma}
n_{i\bar\sigma}\bar n_{j\bar\sigma}=(K_-)^\dagger.
\end{align}
\normalsize
It is easy to verify
that $\sum V_{ij}c^\dagger_{i\sigma}c_{j\sigma}=V(K_0+K_++K_-)$.
Then, using
$[D,K_0]=0$ and $[D,K_+]=\pm K_\pm$ we find:
\begin{equation}
\small
\begin{split}
\bar H_\text{Hub}(t)
&=V(K_0 + e^{iF(t)}K_+ + e^{-iF(t)}K_-) \\
&+(U_0-\Omega)\sum\nolimits_i(n_{i\up}-\tfrac{1}{2})(n_{i\down}-\tfrac{1}{2}).
\end{split}
\end{equation}
The rotated Hamiltonian maintains the $2\pi/\Omega$-periodicity.
Its Fourier components contain the Bessel function
of the first kind
$\int_0^{2\pi}\diff{\tau}
\exp(i(x\cos\tau - m\tau)) = 2\pi i^m\m{J}_m(x)$
through the following expansions:
\small
\begin{gather}
e^{iF(t)} = \sum\nolimits_m
 (-i)^{1+m}\m{J}_{1+m}(\tfrac{\delta U}{\Omega}) e^{-im\Omega t},\\
e^{-iF(t)} = \sum\nolimits_m
 i^{1-m}\m{J}_{1-m}(\tfrac{\delta U}{\Omega}) e^{-im\Omega t}.
\end{gather}
\normalsize
In particular, the average Hamiltonian and the first Fourier components
read:
\begin{subequations}
\small
\label{eq_sm_20}
\begin{align}
\label{eq_sm_2}
\bar H_\text{Hub}^{(0)} &= VK_0-i\m{J}_1(\tfrac{\delta U}{\Omega})VK_+
+i\m{J}_1(\tfrac{\delta U}{\Omega})VK_-\\
\nonumber
&+(U_0-\Omega)
\sum\nolimits_i(n_{i\uparrow}-\tfrac{1}{2})(n_{i\down}-\tfrac{1}{2}),\\
\label{eq_sm_3a}
\bar H_\text{Hub}^{(1)} &= -\m{J}_2(\tfrac{\delta U}{\Omega})VK_+
+\m{J}_0(\tfrac{\delta U}{\Omega})VK_-,\\
\label{eq_sm_3b}
\bar H_\text{Hub}^{(-1)} &= \m{J}_0(\tfrac{\delta U}{\Omega})VK_+
-\m{J}_2(\tfrac{\delta U}{\Omega})VK_-.
\end{align}
\end{subequations}

\subsection{Eqs.~\eqref{eq_4} and \eqref{eq_5} of main text}
Here we calculate the first two terms of the (van-Vleck) large-frequency
expansion, with general expression~\cite{Bukov2016}:
\begin{equation}
\small
H_F = H_0 + \sum_{m>0}\frac{[H_{-m},H_m]}{m\Omega}
+ \m{O}(\tfrac{1}{\Omega^2}).
\end{equation}
The Fourier components $H_m$ are given in
Eqs.~\eqref{eq_sm_20} and depend
on frequency through the Bessel function.
When the large-frequency limit is taken at fixed $\delta U/\Omega$,
this dependence does not show up in the expansion.
In contrast, here we keep $\delta U$ constant and
we have to consider the asymptotic behavior 
$\m{J}_n(\delta U/\Omega) \sim (\delta U/\Omega)^n$.
Then, there are terms in the average Hamiltonian~\eqref{eq_sm_2}
which vanish as $\Omega^{-1}$ and do not enter the
lowest order of the expansion
(cf. Eq.~\eqref{eq_4} of main text):
\begin{equation}
\small
\bar H_\text{Hub}^{\text{eff}(0)}
= VK_0
+(U_0-\Omega)
\sum\nolimits_i(n_{i\uparrow}-\tfrac{1}{2})(n_{i\down}-\tfrac{1}{2}).
\end{equation}
For the same reason, at first order only enter those terms which
actually vanish as $\Omega^{-1}$ 
(cf. Eq.~\eqref{eq_5} of main text):
\begin{equation}
\small
\label{eq_sm_7}
\bar H_\text{Hub}^{\mathrm{eff}(1)} =
(-i\m{J}_{1}(\tfrac{\delta U}{\Omega})VK_+
+ \text{H.c.})
+\tfrac{V^2}{\Omega}(\m{J}_0(\tfrac{\delta U}{\Omega}))^2[K_+,K_-].
\end{equation}

\subsection{Eq.~\eqref{eq_6} of main text}
Here we calculate the commutator in Eq.~\eqref{eq_sm_7}:
\begin{equation}
\small
\label{eq_sm_8}
\begin{split}
[K_+,K_-]
& = \sum_{ij\sigma}\sum_{kl\sigma'}
[c_{i\sigma}^\dagger c_{j\sigma} n_{i\bar\sigma}\bar n_{j\bar\sigma},
c_{k\sigma'}^\dagger c_{l\sigma'} \bar n_{k\bar\sigma'}n_{l\bar\sigma'}] \\
& = \sum_{ij\sigma}\sum_{kl\sigma'}
[c_{i\sigma}^\dagger c_{j\sigma},
c_{k\sigma'}^\dagger c_{l\sigma'}]
n_{i\bar\sigma}\bar n_{j\bar\sigma}\bar n_{k\bar\sigma'}n_{l\bar\sigma'}\\
& + \sum_{ij\sigma}\sum_{kl\sigma'}
c_{k\sigma'}^\dagger c_{l\sigma'}[c_{i\sigma}^\dagger c_{j\sigma},
\bar n_{k\bar\sigma'}n_{l\bar\sigma'}]
n_{i\bar\sigma}\bar n_{j\bar\sigma}\\
& + \sum_{ij\sigma}\sum_{kl\sigma'}
c_{i\sigma}^\dagger c_{j\sigma}[n_{i\bar\sigma}\bar n_{j\bar\sigma},
c_{k\sigma'}^\dagger c_{l\sigma'}]
\bar n_{k\bar\sigma'}n_{l\bar\sigma'}.
\end{split}
\end{equation}
The commutators in Eq.~\eqref{eq_sm_8} give three- and two-site terms.
If we retain only the two-site terms,
then the first sum in Eq.~\eqref{eq_sm_8} reads:
\begin{equation}
\small
\label{eq_sm_12}
\sum\nolimits_{ij}(n_{i\sigma}n_{i\bar\sigma}\bar n_{j\bar\sigma}
-n_{j\sigma}n_{i\bar\sigma}\bar n_{j\bar\sigma})
=\sum\nolimits_{ij}
(n_{i\sigma}-n_{j\sigma})n_{i\bar\sigma}\bar n_{j\bar\sigma}.
\end{equation}
Now, with the identities
$ c_{i\sigma}n_{i\sigma} = c_{i\sigma}$ and
$ n_{i\sigma}c_{i\sigma}  = 0$,
together with their Hermitian conjugates,
it is easy to see that the terms
in the second sum in Eq.~\eqref{eq_sm_8}
are non-vanishing only if $\{i=l,j=k,\sigma'=\bar\sigma\}$, giving:
\begin{equation}
\small
\label{eq_sm_15}
\begin{split}
&\sum\nolimits_{ij} c_{j\sigma}^\dagger c_{i\sigma}
(-c_{i\bar\sigma}^\dagger c_{j\bar\sigma} n_{i\bar\sigma}
- \bar n_{j\bar\sigma}c_{i\bar\sigma}^\dagger c_{j\bar\sigma})\\
&=\sum\nolimits_{ij} c_{i\bar\sigma}^\dagger c_{i\sigma}
c_{j\sigma}^\dagger c_{j\bar\sigma}\\
&=\sum\nolimits_{ij}(c_{i\uparrow}^\dagger c_{i\downarrow}
c_{j\downarrow}^\dagger c_{j\uparrow}
+c_{i\downarrow}^\dagger c_{i\uparrow}
c_{j\uparrow}^\dagger c_{j\downarrow})\\
&=\sum\nolimits_{ij}(S_{i}^+ S_{j}^-+S_{i}^- S_{j}^+)
=2\sum\nolimits_{ij}(S_{i}^x S_{j}^x+S_{i}^y S_{j}^y).
\end{split}
\end{equation}
Analogously, terms in the third sum in Eq.~\eqref{eq_sm_8}
are non-vanishing only if
$\{i=k,j=l,\sigma'=\bar\sigma\}$, giving:
\begin{equation}
\small
\label{eq_sm_13}
\begin{split}
&\sum\nolimits_{ij} c_{i\sigma}^\dagger c_{j\sigma}
(n_{i\bar\sigma}c_{i\bar\sigma}^\dagger c_{j\bar\sigma}
+c_{i\bar\sigma}^\dagger c_{j\bar\sigma}\bar n_{j\bar\sigma})\\
&=\sum\nolimits_{ij} c_{i\sigma}^\dagger c_{j\sigma}
c_{i\bar\sigma}^\dagger c_{j\bar\sigma}
=2\sum\nolimits_{ij} c_{i\uparrow}^\dagger c_{i\downarrow}^\dagger
c_{j\downarrow} c_{j\uparrow}.
\end{split}
\end{equation}

To proceed, we restrict ourselves to a subspace with
double occupancy so large that we can neglect singly occupied sites.
In other words, we restrict ourselves to the local subspace
$\{\ket{0},\ket{\uparrow\downarrow}\}$.
Within this subspace we have
$n_{i\sigma} = n_{i\bar \sigma}$ so that Eq.~\eqref{eq_sm_12} simplifies to:
\begin{equation}
\small
\sum\nolimits_{ij} n_{i\sigma}\bar n_{j\bar\sigma} 
=\sum\nolimits_{ij}
n_{i\sigma}n_{i\bar\sigma}\bar n_{j\bar\sigma}\bar n_{j\sigma}
=2\sum\nolimits_{ij}
n_{i\uparrow}n_{i\downarrow}\bar n_{j\downarrow}\bar n_{j\uparrow}.
\end{equation}
Moreover, within this subspace Eq.~\eqref{eq_sm_15} vanishes, so that the
final result reads (cf. Eq.~\eqref{eq_6} of main text):
\begin{equation}
\small
\label{eq_sm_10}
2\tfrac{V^2}{\Omega}(\m{J}(\tfrac{\delta U}{\Omega}))^2
\sum\nolimits_{ij} 
(c_{i\uparrow}^\dagger c_{i\downarrow}^\dagger
c_{j\downarrow} c_{j\uparrow}+
n_{i\uparrow}n_{i\downarrow}\bar n_{j\downarrow}\bar n_{j\uparrow}).
\end{equation}
This reproduces Eq.(2) of Ref.~\cite{Rosch2008}
for $\delta U=0$, $\Omega=U_0$. 

\subsection{$\eta$-spin ferromagnetic Heisenberg}
To recast Eq.~\eqref{eq_sm_10} to the ferromagnetic Heisenberg model considered
in the main text, we carry out a transformation on the spin-down only:
\small
\begin{align}
\label{eq_sm_31}
c_{i\uparrow}&\rightarrow\tilde c_{i\uparrow} = c_{i\uparrow},\\
c_{i\downarrow}&\rightarrow\tilde c_{i\downarrow}=(-1)^i c_{i\down}^\dagger.
\end{align}
\normalsize
Here $(-1)^{i}=\pm1$ on different sublattices of a bipartite lattice.
Then the first term in Eq.~\eqref{eq_sm_10} transforms to:
\begin{equation}
\small
\begin{split}
-&\sum\nolimits_{ij} (\tilde c_{i\uparrow}^\dagger \tilde c_{i\downarrow}
\tilde c_{j\downarrow}^\dagger \tilde c_{j\uparrow}
+ \tilde c_{i\downarrow}^\dagger \tilde c_{i\uparrow}
\tilde c_{j\uparrow}^\dagger \tilde c_{j\downarrow})\\
=&-\sum\nolimits_{ij}(\eta_{i}^+ \eta_{j}^-+\eta_{i}^- \eta_{j}^+)
=-2\sum\nolimits_{ij}(\eta_{i}^x \eta_{j}^x+\eta_{i}^y \eta_{j}^y).
\end{split}
\end{equation}
This has the same form of Eq.~\eqref{eq_sm_15} with the additional
minus sign $(-1)^{i+j}=-1$ for nearest-neighbor sites.
The $\eta$-spin has the same definition of the physical spin 
for the transformed electrons
$\vec{\eta}_i=\frac{1}{2}\sum_{\alpha\beta}
\tilde c_{i\alpha}^\dagger\vec\sigma_{\alpha\beta}
\tilde c_{i\beta}$ where $\vec\sigma$ is the vector of the three
Pauli matrices:
\begin{equation}
\small
\begin{split}
\eta_i^x &= \frac{\tilde c_{i\uparrow}^\dagger \tilde c_{i\downarrow}
+\tilde c_{i\downarrow}^\dagger \tilde c_{i\uparrow}}{2},\\
\eta_i^y &= \frac{\tilde c_{i\uparrow}^\dagger \tilde c_{i\downarrow}
-\tilde c_{i\downarrow}^\dagger \tilde c_{i\uparrow}}{2i},\\
\eta_i^z &= \frac
{\tilde c_{i\uparrow}^\dagger\tilde c_{i\uparrow}
-\tilde c_{i\downarrow}^\dagger\tilde c_{i\downarrow}}{2}.
\end{split}
\end{equation}
Additionally, if one defines
$\eta_i^+ = \tilde c_{i\uparrow}^\dagger \tilde c_{i\downarrow}$ and 
$\eta_i^- = \tilde c_{i\downarrow}^\dagger \tilde c_{i\uparrow}$
then $\eta_i^x= (\eta_i^++\eta_i^-)/2$ and
$\eta_i^y=(\eta_i^+-\eta_i^-)/2$.

To consider the second term in Eq.~\eqref{eq_sm_10}, we notice
that under the transformation~\eqref{eq_sm_31} the density operator
transforms as
$n_{i\uparrow}\rightarrow\tilde n_{i\uparrow} = n_{i\uparrow}$ and 
$n_{i\downarrow}\rightarrow\tilde n_{i\downarrow}=1- n_{i\down}$.
Then, this term transforms to
(keep in mind that in the considered subspace
$n_{i\sigma} = n_{i\bar \sigma}$):
\begin{equation}
\small
2\sum\nolimits_{ij} \tilde n_{i\bar\sigma} \tilde n_{j\sigma} 
=\sum\nolimits_{ij}(\tilde n_{i\uparrow}\tilde n_{j\downarrow}+
\tilde n_{i\downarrow}\tilde n_{j\uparrow}) 
=-2 \sum\nolimits_{ij} (\eta_i^z\eta_j^z-\tfrac{1}{4}).
\end{equation}
The last equality is best demonstrated verifying that the operators
have the same matrix elements
in the considered subspace.
Alternatively, an explicit derivation reads:
\begin{equation*}
\small
\begin{split}
&\tilde n_{i\uparrow}\tilde n_{j\downarrow}
+\tilde n_{i\downarrow}\tilde n_{j\uparrow}\\
&=\tfrac{1}{2}(\tilde n_{i\uparrow}\tilde n_{j\downarrow}
+\tilde n_{i\downarrow}\tilde n_{j\uparrow})
+\tfrac{1}{2}(\tilde n_{i\uparrow}\tilde n_{j\downarrow}
+\tilde n_{i\downarrow}\tilde n_{j\uparrow})\\
&=\tfrac{1}{2}(\tilde n_{i\uparrow}\tilde n_{j\downarrow}
+\tilde n_{i\downarrow}\tilde n_{j\uparrow})
+\tfrac{1}{2}(\tilde n_{i\uparrow}(1-\tilde n_{j\uparrow})
+\tilde n_{i\downarrow}(1-\tilde n_{j\downarrow}))\\
&=-2(\tfrac{1}{4}(\tilde n_{i\uparrow}
-\tilde n_{i\downarrow})(\tilde n_{j\uparrow}-\tilde n_{j\downarrow})
-\tfrac{1}{4}).
\end{split}
\end{equation*}
Here it is crucial the use of
$\tilde n_{i\uparrow}+\tilde n_{i\downarrow}=1$.

\section*{VI. Role of drive amplitude}
\begin{figure}
\includegraphics[width=\columnwidth]{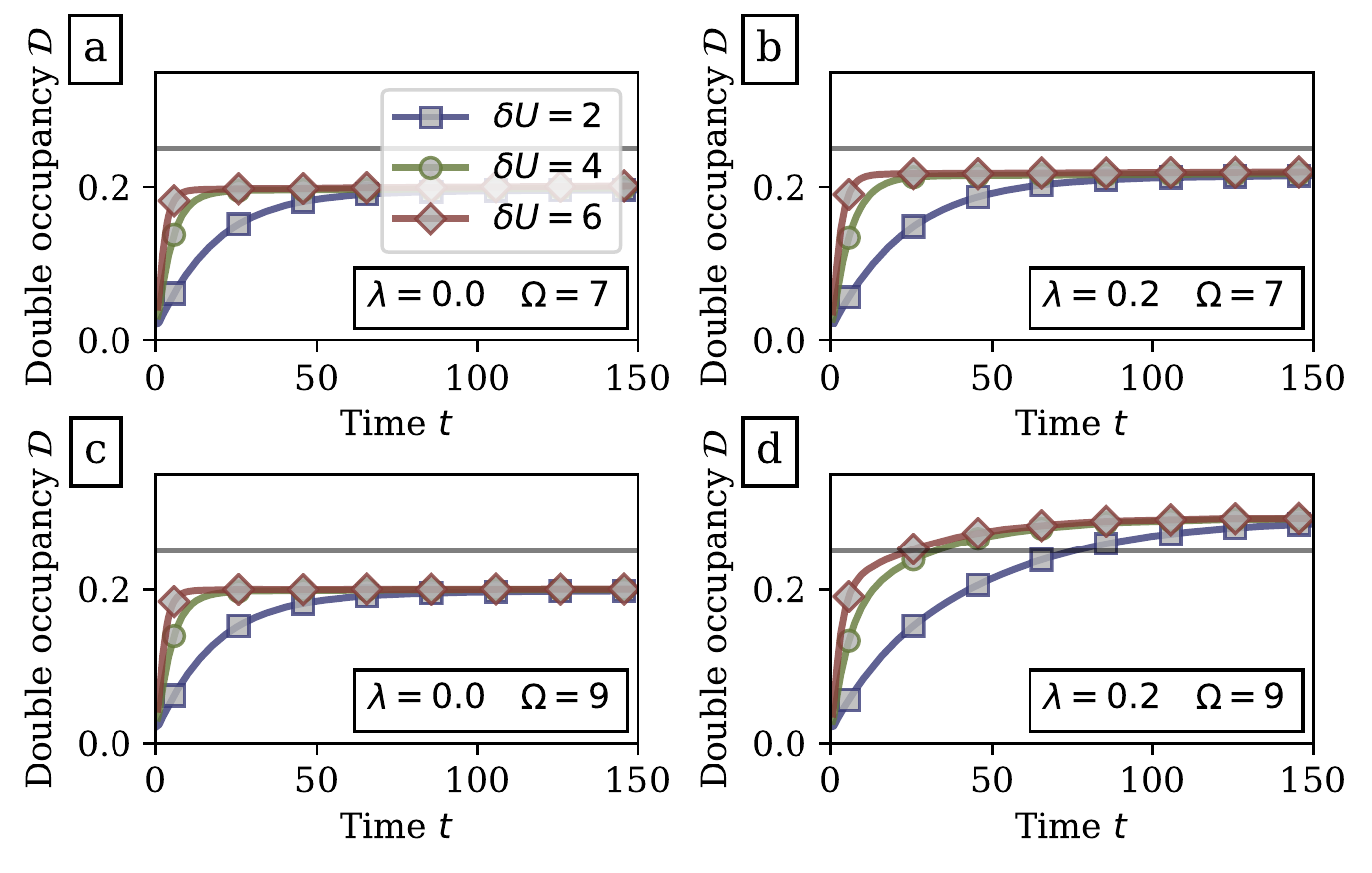}
\caption{Time evolution of double occupancy
for the isolated model ($\lambda=0.0$) and the dissipative one ($\lambda=0.2$),
for drive frequencies below ($\Omega=7$) and above ($\Omega=9$) resonance,
and for different drive amplitude $\delta U=2,4,6$.
}
\label{fig_sm_7}
\end{figure}

In Fig.~\ref{fig_sm_7} we compare the time evolutions for
various drive amplitudes $\delta U$. These show interesting features
in both the isolated model (see also Ref.~\cite{Peronaci2018})
and the dissipative model.
First, it is evident how the drive amplitude controls the time scale of the
transient dynamics, with larger amplitudes inducing shorter transients.
Second, the steady steady, in sharp contrast, appears to be largely independent
from the drive amplitude.

These observations are consistent with 
and provide further validation to our analysis based on
the Floquet theory (Eqs.~(3) to (5) of main text). Indeed, in Eq.~(4)
a factor $\m{J}_1(\delta U/\Omega)\sim\delta U/\Omega$
multiplies the terms creating doublon excitations, which
are responsible for the increase of double occupancy
in the transient dynamics. Thus, a large drive amplitude enhances
these terms and makes the transient shorter.
Moreover, since the long-time values do not depend on $\delta U$,
it is confirmed that these terms
are largely suppressed in the steady state, which validates the introduction
of the doublon-only Hamiltonian (5).

\section*{VII. Hypercubic lattice}
\begin{figure}
\includegraphics[width=\columnwidth]{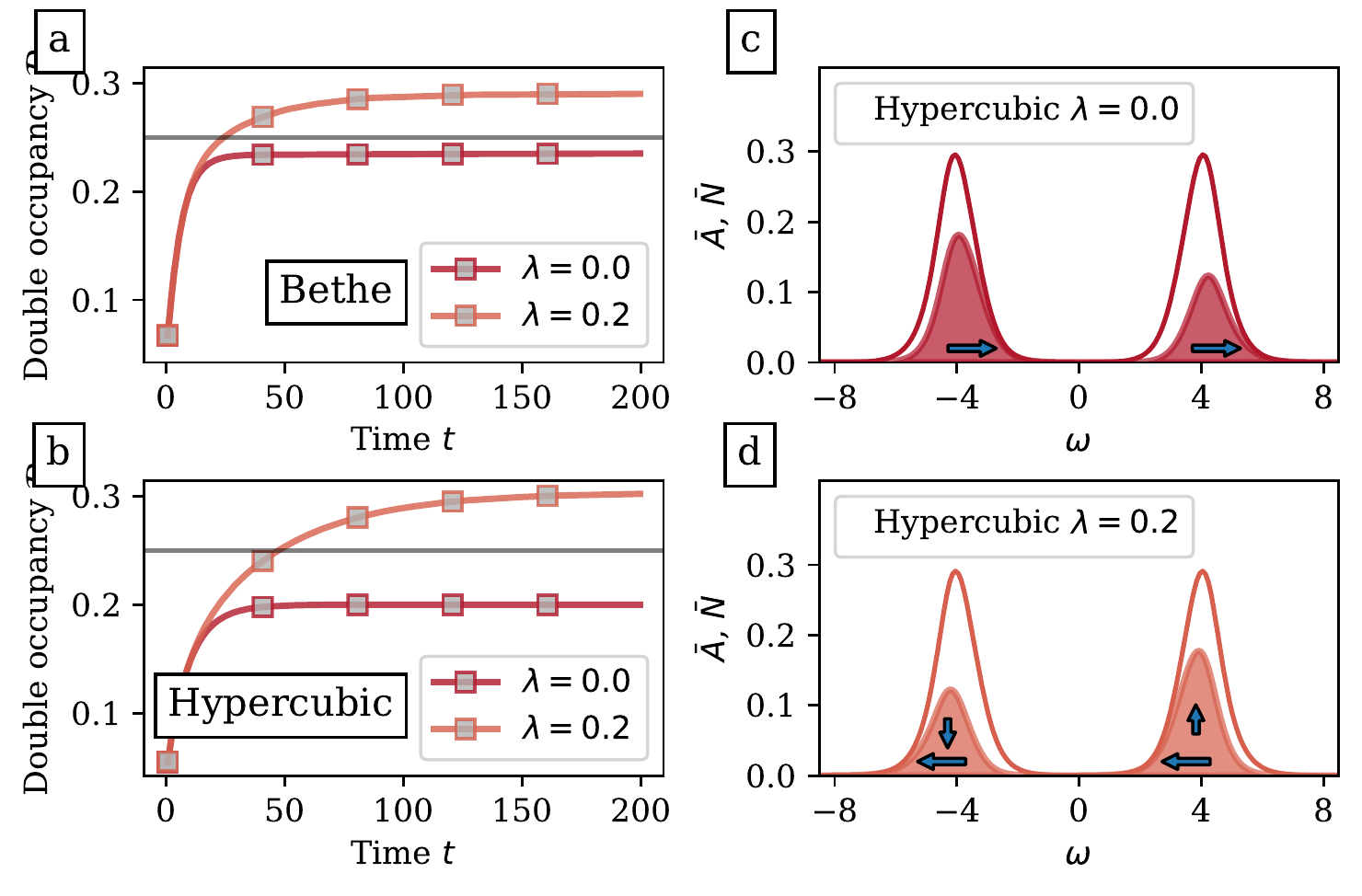}
\caption{Time evolution of double occupancy
for drive frequency $\Omega=9>\Omega^*\simeq U_0$ and amplitude
$\delta U=4$ on Bethe (a) and hypercubic (b) lattices for
$\lambda=0.0$ (isolated) and $\lambda=0.2$ (dissipative).
Corresponding long-time average spectral function
$\bar A(\omega)$ (solid line) and occupation function $\bar N(\omega)$
(filled area) for the hypercubic lattice, showing
population inversion in the isolated model (c),
signaled by the blueshift of $\bar N(\omega)$ with
respect to $\bar A(\omega)$ (arrows); and large double occupancy in the
dissipative model (d), signaled by the increase of $\bar N(\omega)$
in the high-energy band (vertical arrows).}
\label{fig_sm_8}
\end{figure}

In the main text we consider a system on Bethe lattice,
whose free-electron density of states is semicircular.
Here we consider a hypercubic lattice in infinite dimensions, whose
free-electron density of states is Gaussian.
In Fig.~\ref{fig_sm_8} we show that this difference 
does not qualitatively affect the physics discussed in the main
text. For drive frequency above resonance $\Omega=9>\Omega^*\simeq U_0$
we observe qualitatively the same increase in the long-time double occupancy
which goes above $0.25$ in the dissipative model ($\lambda=0.2$) both
on Bethe lattice (Fig.~\ref{fig_sm_8}a) and on hypercubic lattice
(Fig.~\ref{fig_sm_8}b).

The spectral and occupation function on hypercubic lattice also show the same
changes between the isolated (Fig.~\ref{fig_sm_8}c) and the dissipative
(Fig.~\ref{fig_sm_8}d) models. In the former, we observe population inversion
within each of the Hubbard bands. In the latter, as a consequence of
dissipation, the population inversion within the Hubbard bands does not take
place. At the same time, the occupation of the upper Hubbard band grows larger
than the one of the lower Hubbard band (cf. Fig.~2 of main text).

\section*{VIII. Bath spectral function}
\begin{figure}
\includegraphics[width=\columnwidth]{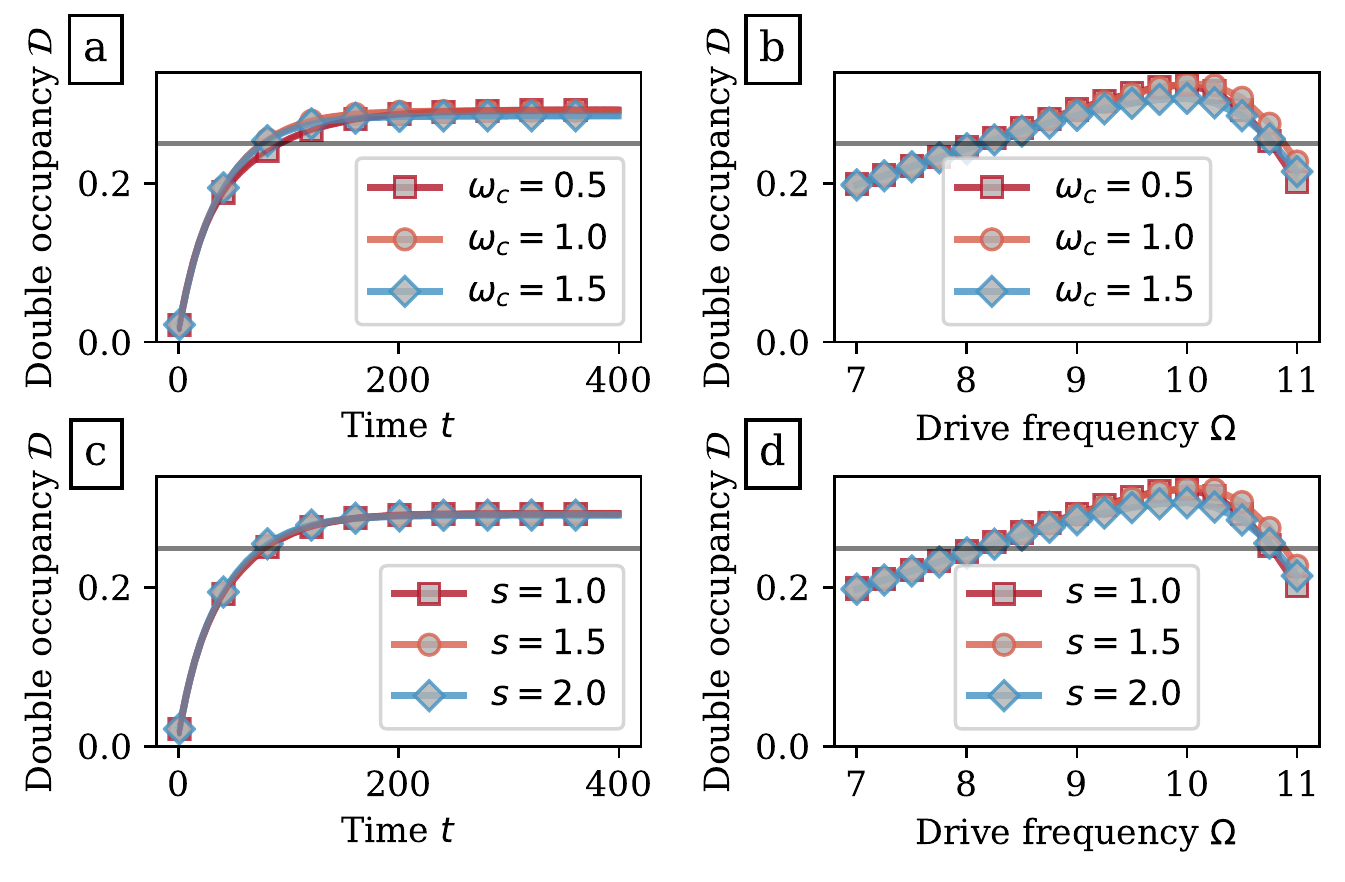}
\caption{Time evolution for drive frequency 
$\Omega=9>\Omega^*$ and long-time average of
double occupancy as a function of drive frequency for 
various bath parameter $\omega_c=0.5,1,1.5$ at fixed $s=2$ (a-b)
and $s=1,1.5,2$ at fixed $\omega_c=1$ (c-d).}
\label{fig_sm_9}
\end{figure}

In the main text we consider a bosonic thermal bath with spectral function
$J(\omega) =  \frac{\omega_c}{2} (\frac{\omega}{\omega_c})^s
e^{-\frac{\omega}{\omega_c}}$ with $\omega_c=1$ and $s=2$.
Here in Fig.~\ref{fig_sm_9} we provide additional numerical results with
various parameters $\omega_c$ and $s$.

\end{document}